# Chemical Aspects of the Antiferromagnetic Topological Insulator MnBi$_2$Te$_4$

Alexander Zeugner[1], Frederik Nietschke[2], Anja U. B. Wolter[3], Sebastian Gaß[3], Raphael C. Vidal[4], Thiago R. F. Peixoto[4], Darius Pohl[3,5], Christine Damm[3], Axel Lubk[3], Richard Hentrich[3], Simon K. Moser[4,6], Celso Fornari[4], Chul Hee Min[4], Sonja Schatz[4], Katharina Kißner[4], Max Ünzelmann[4], Martin Kaiser[1], Francesco Scaravaggi[3], Bernd Rellinghaus[3,5], Kornelius Nielsch[3,7,8], Christian Heß[3], Bernd Büchner[3,9], Friedrich Reinert[4], Hendrik Bentmann[4], Oliver Oeckler[2], Thomas Doert[1], Michael Ruck[1,10], Anna Isaeva[1*]

1 Faculty of Chemistry and Food Chemistry, Technische Universität Dresden, 01062 Dresden, Germany

2 Institute for Mineralogy, Crystallography and Materials Science, Leipzig University, 04275 Leipzig, Germany

3 Leibniz-Institute for Solid State and Materials Research, 01069 Dresden, Germany

4 Experimentelle Physik VII, Universität Würzburg, 97074 Würzburg, Germany

5 Dresden Center for Nanoanalysis, cfaed, Technische Universität Dresden, 01062 Dresden, Germany

6 Advanced Light Source, Lawrence Berkeley National Laboratory, Berkeley, CA 94720, USA

7 Institute of Materials Science, Technische Universität Dresden, 01062 Dresden, Germany

8 Institute of Applied Physics, Technische Universität Dresden, 01062 Dresden, Germany

9 Institute of Solid State Physics, Technische Universität Dresden, 01062 Dresden, Germany

10 Max Planck Institute for Chemical Physics of Solids, 01187 Dresden, Germany

**ABSTRACT:** Crystal growth of MnBi$_2$Te$_4$ has delivered the first experimental corroboration of the 3D antiferromagnetic topological insulator state. Our present results confirm that the synthesis of MnBi$_2$Te$_4$ can be scaled-up and strengthen it as a promising experimental platform for studies of a crossover between magnetic ordering and non-trivial topology. High-quality single crystals of MnBi$_2$Te$_4$ are grown by slow cooling within a narrow range between the melting points of Bi$_2$Te$_3$ (586 °C) and MnBi$_2$Te$_4$ (600 °C). Single crystal X-ray diffraction and electron microscopy reveal ubiquitous antisite defects in both cation sites and, possibly, Mn vacancies. Powders of MnBi$_2$Te$_4$ can be obtained at subsolidus temperatures, and a complementary thermochemical study establishes a limited high-temperature range of phase stability. Nevertheless, quenched powders are stable at room temperature and exhibit long-range antiferromagnetic ordering below 24 K. The expected Mn(II) out-of-plane magnetic state is confirmed by the magnetization, X-ray photoemission, X-ray absorption and linear dichroism data. MnBi$_2$Te$_4$ exhibits a metallic type of resistivity in the range 4.5–300 K. The compound is an $n$-type conductor that reaches a thermoelectric figure of merit up to $ZT$ = 0.17. Angle-resolved photoemission experiments provide evidence for a surface state forming a gapped Dirac cone.

## INTRODUCTION

A sizeable portion of the present research on topological materials[1] is conducted in pursuit of candidate materials for the quantum anomalous Hall effect (QAHE)[2,3] and the topological magnetoelectric effect.[4] One of the most compelling prospects of topological phases is the manifestation of quantum effects under ambient conditions. A magnetic topological insulator (TI) with large spin-orbit coupling could exhibit quantized resistance and, concurrently, one non-dissipative spin-polarized channel along the edges at room temperature (RT) without an external magnetic field.[5]

The QAHE was realized first in 2013 in a thin film of the Cr- and Bi-doped topological insulator Sb$_{2-x}$Te$_3$ in the mK range.[6] Moreover, high external magnetic fields above 10 T were required to suppress the longitudinal resistance (dissipative conduction channels).[6] V-doped samples[7] showed superior properties compared to Cr, but, nevertheless, the signatures of the QAH state vanished above ~5 K. Promising results have been obtained for magnetically modulation-doped topological insulators, where the QAHE stabilizes up to 1–2 K.[8–10] Alternative QAHE proposals are based on the proximity effect, i. e. on the direct coupling via a shared interface between a bulk (anti)ferromagnet and a TI.[11–17] Theory has demonstrated some disadvantages of this approach[18] and has introduced a new principle coined "magnetic extension".[19,20] According to this concept, the structural similarity between a magnetically ordered overlay and a TI substrate promotes better overlap between their states, in particular, in layered materials. A perspective candidate magnetic TI, thus, should have a crystal structure resembling those of known topological insulators. Magnetic doping (Mn, Fe, Cr, V) in the classic topological insulators, above all in $Bi_2Te_3$, is faced with challenges.[21] A material with a *periodic crystal structure* and an *intrinsic magnetic ordering* of metal atoms ordered in distinct crystallographic positions would be a highly desired alternative. An added asset would be van der Waals gaps that ensure a gradual change in the chemical potential at the interface and a less problematic experimental realization of heterostructures and devices.

The unique compound that fulfills these prerequisites, $MnBi_2Te_4$, has been put forward and scrutinized by theory[20, 22–24] for some time after its discovery.[25] The latter work has presented the synthesis of polycrystalline samples and a crystal-structure determination by the Rietveld method, thermoelectric properties of a composite material obtained by decomposition and the calculated density of states. Ensuing experimental progress was limited by absence of single crystals and robust synthesis protocols.

$MnBi_2Te_4$ is a layered compound built up by septuple layers $^2_\infty$[Te–Bi–Te–Mn–Te–Bi–Te] stacked in the rhombohedral (*ABC*) fashion. It has to be noted that compounds offering a combination of bismuth (a bearer of strong spin-orbit coupling as an essential ingredient for a topological band inversion[1]) and 3*d* transition metals (cooperative magnetic phenomena) are extremely sparse. Direct bonding is avoided between Bi and most of the 3*d* metals (Cr, V, Mn, Fe, Co, Cu), and the respective binary phase diagrams show either a complete lack of miscibility[26,27] or feeble solubility.[28] MnBi, a ferromagnet with a high magnetocrystalline anisotropy,[29] is difficult to synthesize due to omnipresent phase separation. A handful of intermetallic high-pressure phases, e. g. $FeBi_2$, CuBi, $Cu_{11}Bi_7$, β-NiBi,[30] and $CoBi_3$,[31] constitute chemical rarities.

Our recent joint theory–experiment study[32] establishes $MnBi_2Te_4$ as the first 3D antiferromagnetic topological insulator (AFM TI) since a prediction of this state in 2010.[33] The topological character of the band structure has been demonstrated by first-principles calculations, while the experimental corroboration of the intrinsic antiferromagnetic ordering and a gapped surface state has been obtained on $MnBi_2Te_4$ single crystals grown for the first time. Follow-up manuscripts[34–36] support these claims and take a further step of proposing experimental set-ups, in which signatures of proximity-induced Majorana modes and quantized magnetoelectric effect could be observed in a real material $MnBi_2Te_4$. These prospects of functionalization generate a high demand on high-quality well-characterized samples.

To address this boosting interest, we herewith report a full account on our developed crystal growth technique along with the structure refinement, powder synthesis, and a phase-stability study of $MnBi_2Te_4$. Physical properties elucidated on polycrystalline samples support and extend the results obtained on single crystals.[32] We show that, despite the material's peculiar thermal behavior, synthesis of $MnBi_2Te_4$ can be scaled up for various studies and applications.

## RESULTS AND DISCUSSION

**1. Powder Synthesis and Thermodynamic Stability.** $MnBi_2Te_4$ was reported to be metastable[25] based on the following observations. A $MnBi_2Te_4$ powder sample was obtained after annealing of a fused, quenched ingot at 535 °C for 48 hours. It had decomposed into presumably non-stoichiometric $Bi_2Te_3$ and $MnTe_2$ during an in situ high-temperature powder X-ray diffraction experiment, where the decomposition of the powder started at 150 °C and was complete at 500 °C.

Following the procedure in Ref. [25], we have synthesized a sample (#1) from a stoichiometric mixture of the elements fused at 950 °C for 1 day, air-quenched, annealed at ca. 580 °C for 9 days and air-quenched again. The polycrystalline product contained primarily $MnBi_2Te_4$ and a small admixture of MnTe (Figure S1, approx. 4 wt.% according to a Rietveld refinement).

Another sample (#2) was obtained via a solid-state reaction from a stoichiometric mixture of pre-synthesized precursors, α-MnTe (NiAs-type[37]) and $Bi_2Te_3$ (Figure S2). The sample was water-quenched. Prior to the annealing, the mixture was homogenized via ball milling that, however, did not yield ternary products (Figure S3). Annealing at 565 °C for 10 days afforded a nearly phase-pure ternary product that contained that about 2 wt.% of MnTe (Figure 1), similar to sample #1.

The similar outcome of these two experiments suggests that $MnBi_2Te_4$ may be thermodynamically stable in a limited temperature interval and may deviate from the idealized stoichiometric composition. Subsequently, the thermal stability of the synthesized $MnBi_2Te_4$ powders was studied by a combination of differential scanning calorimetry (DSC), long-term annealing at various temperatures, and temperature-programmed powder diffraction experiments (Figure 1, see also SI for a detailed discussion, Figure S4–S10, Table S1).

From the DSC experiments performed on a stoichiometric mixture of binary precursors (Figure S4) and on presynthesized $MnBi_2Te_4$ samples (Figure 1 for sample #2, Figure S7 for sample #1), we conclude that $MnBi_2Te_4$ melts at ca. 600 °C ($T_{ons}$ = 595 °C, $T_{peak}$ = 599 °C). The exact character of melting could not be determined. $MnBi_2Te_4$ presumably melts incongruently since solidification of either $Bi_2Te_3$ or even some hitherto unknown ternary Mn-containing phase has been observed in the DSC cooling curves of all our samples (cf. SI). At subsolidus temperatures, the DSC data did not signal further decomposition of $MnBi_2Te_4$ (Figure S6), neither did they reflect the formation of $MnBi_2Te_4$ upon heating.

In contrast to that, long-term annealing of the presynthesized MnBi$_2$Te$_4$ powders (sample #2) at 400 °C and below results in their inevitable decomposition into Bi$_2$Te$_3$ and MnTe (Figure 1). Decomposition takes days and proceeds notably slower at 200 °C than at 400 °C. Nevertheless, this process is reversible as the obtained phase mixtures can recombine into MnBi$_2$Te$_4$ if the above-described synthesis procedures are repeated. The decomposition rate appears to depend on the particle size: finely dispersed powders decompose faster at lower temperatures than coarse powders or compacted bulk with the same phase composition.

As opposed to the DSC data (Figure S6), temperature-programmed powder diffraction on presynthesized MnBi$_2$Te$_4$ (sample #1) up to ca. 650 °C concerted with the results in Ref. [25]. At a heating rate of 2 K min$^{-1}$, decomposition of MnBi$_2$Te$_4$ into Bi$_2$Te$_3$ and, presumably, manganese tellurides has commenced already at ca. 200 °C (Figure 1) and was faster in finely ground powders. Upon cooling, MnBi$_2$Te$_4$ did not form, unless the mixture was fused again, repeating the original synthesis protocol. This might be explained by rather macroscopic phase separation. In a similar way, the phase composition and crystallinity of larger bulk samples depend on the quenching/cooling rates. The fraction of MnBi$_2$Te$_4$ significantly increased at sufficiently slow cooling rates (6 K h$^{-1}$) (Figure S10).

These seemingly contradictory results point at strong kinetic hindrances in the formation and decomposition of MnBi$_2$Te$_4$. An increased mobility in the melt may be a beneficial factor for the formation of MnBi$_2$Te$_4$. To conclude the discussion of phase stability, we put forward an assumption that MnBi$_2$Te$_4$ is thermodynamically stable in a rather narrow high-temperature interval below 600 °C, and its decomposition in compact bulk may be hindered due to an interplay of diffusion and strain. The formation temperature lies somewhere between 400 °C (our data) and 535 °C,[25] but remains to be determined. Owing to this kinetic barrier, powder crystalline samples of MnBi$_2$Te$_4$, which were synthesized within the phase stability interval, can be quenched down to room temperature, where MnBi$_2$Te$_4$ is metastable. Compacted bulk samples and crystals can be used for physical property measurements (*vide infra*) at low and slightly elevated temperatures without showing signs of massive decomposition.

The issue of phase purity is intimately related to this point. All powder samples contained minor side phases, typically MnTe, that were detected by PXRD and/or by EDX. These ubiquitous admixtures could stem from partial decomposition upon quenching, as just discussed above, or from an incomplete reaction of the starting materials. MnTe remains solid when Bi$_2$Te$_3$ is already molten, so the reaction is always heterogeneous. Non-stoichiometry of the title compound is yet another possible explanation. An additional signal in the DSC heating curve of sample #2 (Figure 1) could indicate an eutectic between MnTe and the ternary phase. Following this argument, decomposition of the latter could be associated with phase relations that do not involve the ternary phase. Moreover, single-crystal X-ray diffraction (SCXRD) experiments establish that the title compound could tolerate a significant degree of non-stoichiometry (*vide infra*). However, experiments targeting a non-stoichiometric variant Mn$_{0.85}$Bi$_{2.1}$Te$_4$ by both synthesis routes #1 and #2 have also yielded a nearly phase-pure sample according to PXRD (Figure S11). The thermal behavior of these samples in DSC experiments was similar to that of the previously studied samples (Figure S12 and the discussion in the SI). So we conclude that phase purity of MnBi$_2$Te$_4$ samples is mainly determined by its finite temperature-stability window.

**2. Crystal Growth Optimization.** Based on the results above, we developed a crystal-growth route for MnBi$_2$Te$_4$. The small energetic difference between the formation enthalpies of Bi$_2$Te$_3$ and MnBi$_2$Te$_4$ is indicated by a gap of only 9 K between the melting temperatures of Bi$_2$Te$_3$ and MnBi$_2$Te$_4$ (and less than 6 K in the solidification temperatures due to different undercooling). These findings suggest that single-crystal growth can proceed via slow cooling over the very narrow Ostwald–Miers region between the onset temperatures of melting (signal **T$_2$**) and solidification (signal **T$_3$**) (cf. Figure 1). Indeed, single crystals were obtained by slow cooling in the Ostwald–Miers region (ca. 3 K wide) followed by annealing for several days. All samples were water-quenched. Platelet-shaped crystals with diameters up to 200 μm were obtained atop of a highly crystalline ingot. Our PXRD analysis of both crystals and the ingot revealed nearly phase-pure MnBi$_2$Te$_4$ (Figure 2) with a strongly preferred orientation along the [001] direction. It should be noted that the polycrystalline samples obtained in this way were never absolutely phase pure as indicated by slight asymmetry of some reflections in Figure 2.

**3. Crystal Structure Refinement.** We were now able to elucidate the crystal structure of MnBi$_2$Te$_4$ by SCXRD. In agreement with previously reported powder data,[25] MnBi$_2$Te$_4$ crystallizes in the centrosymmetric space group $R\bar{3}m$ (no. 166) with the lattice parameters $a$ = 4.3283(2) Å and $c$ = 40.912(3) Å. Electron diffraction patterns collected on a wedge of a thin lamella cut out from a single crystal agree well with the trigonal symmetry and the unit cell found in SCXRD (Figure 2).

The reported crystal structure of MnBi$_2$Te$_4$[25] belongs to the ordered so-called "GeAs$_2$Te$_4$" structure type, which comprises septuple layers. However, the experimental electron density did not correspond to two specific fully occupied Mn and Bi positions. Cation intermixing is ubiquitous in the abundant mixed $Tt$Bi$_2$Te$_4$, $Tt$ = Ge, Pb, Sn semiconductors.[38] Moreover, reported ordered models have only been postulated, while structure refinements have always yielded cation disorder, although compounds of this structure type typically do not contain 3$d$ metals with an exception of rare instances of doping.[39]

For this reason, cation intermixing in the Bi and Mn sites in MnBi$_2$Te$_4$ was taken into account. We considered antisite defects (Bi in the Mn site and *vice versa*) and cation vacancies (Figure S13, Table 1, 2, S2–S7). Charge neutrality of the title compound was assumed in the structure refinements under the restraint of 8 positive charges per formula unit. The oxidation states in the stoichiometric variant can be directly ascribed to $Mn^{II}(Bi^{III})_2(Te^{-II})_4$, similar to structurally related compounds like $GeBi_2Te_4$.[40] This also seems realistic from a chemical viewpoint since tellurium is not expected to oxidize manganese to its higher oxidation states.

The resultant non-stoichiometric $Mn_{0.85(3)}Bi_{2.10(3)}Te_4$ model comprises 21.5(1)% Bi, 73.6(1)% Mn and 4.9(1)% voids on the 3a position, whereas the Bi site (6c) has 5.7(1)% of Mn antisite defects (Figure 2). Vacancies were allowed only in the 3a position; they were considered less probable for the 6c site, since the Te atoms border a van der Waals gap and already have only three bonding partners.

Cation intermixing is a structural hallmark — a hypothetical ordered structure model yields a non-stable refinement. Since the presence of vacancies is less significant, a second model without vacancies but still with mixed occupancies of both cation positions was considered (Table S4–S7). However, it yielded only slightly higher R-factors. In this scenario, the Mn site has 17.5(1)% of Bi antisite defects, while the Bi site (with twice the multiplicity of the Mn site) hosts 8.8(1)% Mn defects. The restraint of the charge neutrality was also applied in this solution, so that the sum formula is $MnBi_2Te_4$.

In order to further elaborate the details of the crystal structure, HR-STEM imaging and local elemental mapping using electron energy loss spectroscopy (EELS) was used. Figure 2 summarizes the results in the [100]* zone axis. Even though the investigation was conducted at 80 kV acceleration voltage only, sample damage due to the electron beam could not be avoided completely, and made the analysis challenging. As can be seen from the intensity of the HAADF (high angle annular dark field) image, the single atomic planes vary in their intensity distribution. The intensity scales approximately linearly with the line integral of the atomic number that are each raised to a power of about 1.7. The strong intensity in the central atomic columns of a septuple layer is surprising, as this position is expected to be mainly occupied by Mn atoms and only by a small fraction of Bi atoms. Using local elemental mapping by analyzing EELS spectra of the scanned area, Mn and Te distributions have been visualized. It is found, that Mn is distributed over the septuple layer instead of being localized solely in its central atomic plane; thus, corroborating the results of X-ray diffraction.

Based on the available results, we cannot make a conclusive choice for one of the structure models. Summing up, it seems likely that the structure type tolerates some variations in the site occupancies. In an attempt to tackle this issue with EDX, spectra of several crystals (averaged over ca. 10 points for each) were quantified using stoichiometric $Bi_2Te_3$ and MnTe as references. Averaging of the obtained data over the different choices of the reference compound (e. g. for Te) and edges (Bi-M or Bi-L) used for quantification has resulted in the composition (in at. %): Mn 11.5(9), Bi 31(1), and Te 58(2), corresponding to $Mn_{0.81(6)}Bi_{2.1(2)}Te_4$. The rather large uncertainties in the quantification of the heavy element concentration stem likely from systematic errors introduced by different edges and reference choices and do not allow us to distinguish between the two structure models, $Mn_{0.85(3)}Bi_{2.10(3)}Te_4$ and $MnBi_2Te_4$.

Our powder synthesis yielded nearly phase-pure samples for both compositions (vide supra). Thus, solid solutions $Mn_{1-x}Bi_{2+2x/3}Te_4$, which necessitate charge neutrality, assuming Mn(II), Bi(III) and Te(−II), appear to be likely. This is not unusual with respect to related compounds $(GeTe)_nSb_2Te_3$: they do not always form the structure type that is initially suggested by the composition.[41] For the sake of convenience, we continue to use the sum formula $MnBi_2Te_4$ instead of $Mn_{1-x}Bi_{2+2x/3}Te_4$ throughout the following text.

An interesting structural aspect is that $MnBi_2Te_4$ strongly resembles the classic TI, $Bi_2Te_3$. The septuple layers of the former may be regarded as a flat arrangement of edge-sharing Mn-centered telluride octahedra incorporated into the quintuple layers of $Bi_2Te_3$. Once again, Bi and Mn do not engage into direct bonding, as they occupy the octahedral voids in a close-packed array of Te atoms and, thus, reside at the nearest distances of 4.33 or 4.51 Å. The coordination of the Bi atoms is asymmetric: the Bi–Te distances towards the Mn layer are notably longer than the Bi–Te towards the van der Waals gap (Figure 2). Thus, $MnBi_2Te_4$ could be regarded as $(BiTe)^+(MnTe_2)^{2-}(BiTe^+)$ in a semi-ionic model, similar to $(BiTe)^+I^-$.[42] These crystal-chemical similarities may inspire the design of modular structures between the title compound and $Bi_nTeX$, X = I, Br topological materials.[43,44] The structural compatibility may enable a strong overlap between the topological surface states of $Bi_2Te_3$ or $Bi_nTeX$ and the spin-polarized interface states of $MnBi_2Te_4$. The first example has already been considered theoretically in a model of a septuple $^2_\infty$[Te–Bi–Te–Mn–Te–Bi–Te] overlay atop of semi-finite $Bi_2Te_3$ bulk.[19]

**4. Spectroscopic Characterization.** The electronic and magnetic properties of $MnBi_2Te_4(001)$ single-crystals were studied by X-ray photoelectron spectroscopy (XPS) and X-ray absorption spectroscopy (XAS). The XAS line shape of the Mn $L_{2,3}$ absorption edge (Figure 3) confirms the $Mn^{II}$ valence state, based on a comparison to reference data on different Mn oxides and fluorides, namely MnO, $Mn_2O_3$, $MnO_2$, $MnF_2$ and $MnF_3$.[45,46] Similarly, the XPS line shapes of the Mn 2p levels (Figure 3) resemble spectra of Mn compounds in which Mn is divalent.[47, 48] In particular, the satellites, located approximately 6 eV below the two main lines, are characteristic of $Mn^{II}$.[49] Both XPS and XAS show no evidence for the presence of two different Mn species, despite two possible Mn atomic positions which were established by XRD and EELS in the previous section. We therefore expect the chemical states of the Mn ions on the 6c and 3a sites to be similar and, thus, without considerable differences in the XAS/XPS line shapes and positions.

The band structure of $MnBi_2Te_4(001)$ near the Fermi level has been examined by angle-resolved photoelectron spectroscopy (ARPES). We observe an upper electron-like feature and a lower hole-like feature which are separated by an energy gap at the $\bar{\Gamma}$ point ($k_{\parallel} = 0$) of approximately 100 meV (Figure 3). The ARPES data agree qualitatively with previous *ab initio* calculations for MnBi$_2$Te$_4$(001).[32] Based on a comparison with these calculations, we interpret the measured bands in Figure 3 as a gapped Dirac cone, which is indeed expected for a topological insulator with intrinsic out-of-plane magnetization. Furthermore, the ARPES measurements indicate an *n*-doped character of MnBi$_2$Te$_4$, in contrast to the *p*-type semiconductor in Ref. [25].

X-ray linear dichroism (XLD) data collected in total electron yield (TEY) mode (Figure 3) confirm the previous magnetic characterization.[32] Measurements performed in grazing light incidence geometry at $T = 2$ K show a clear XLD signal, indicative of an out-of-plane anisotropy of the Mn $3d$ charge distribution. At $T = 40$ K, i. e. above $T_N$, no XLD signal is detected anymore, which confirms the magnetic origin of the XLD signal at low $T$, as opposed to a crystal field effect. The out-of-plane nature of the magnetic state is further supported by the absence of a XLD signal at $T = 2$ K in normal light incidence geometry. The probing depth of the XLD experiment is on the order of only several nm, which implies that the out-of-plane magnetic state, deduced from the bulk sensitive SQUID experiments,[32] extends towards the surface.

**5. Magnetic and Transport Properties.** According to DFT+$U$ calculations,[22,32] the ground state of the ideally ordered bulk MnBi$_2$Te$_4$ corresponds to an interlayer antiferromagnet with ferromagnetic intralayer coupling and an inverted topological band gap of ca. 200 meV. Experimental evidence of the long-range antiferromagnetic ordering below $T_N = 24$ K and an energy gap of 70 meV has been delivered on our single crystals with the above described Bi/Mn intermixing.[32]

The magnetic characteristics of a polycrystalline melt ingot (part of sample #1) are presented in this work and are in good accordance with previous results on our single crystals (Figure 4). We do not find any additional magnetic contribution associated with Mn antisite defects in the Bi sites. Apart from the clear long-range AFM order at low temperatures ($T_N = 24.6(5)$ K), the paramagnetic regime (150–300 K) resembles a Curie-Weiss law, $\chi(T) = \chi_0 + C/(T-\theta_{CW})$. Here $\chi_0$ stands for a small temperature independent magnetic susceptibility of both diamagnetic closed electron shells and a Pauli paramagnetic contribution due to some metallicity in this material. The fitted values for the effective moment and the Curie–Weiss temperature, $\mu_{eff} = 5.4(1)\mu_B$ and $\theta_{CW} = +1(3)$ K, are in very good agreement with our work on single crystals.[50] The $M(H)$ curve at $2$ K $< T_N$ further shows an indicative spin-flop transition at $\mu_0 H_{SF} = 3.7(2)$ T, in line with Ref. [32].

Our temperature-dependent specific heat measurements on an assembly of several single crystals (Figure 4) demonstrate a non-singular increase at $T_N \sim 24.2(2)$ K accompanying the antiferromagnetic ordering that is characteristic for a second-order phase transition. It is broadened and nearly suppressed in fields $H > H_c$ (e. g. 9 T) applied perpendicular to (001).

Complementary to the magnetic properties, the temperature dependence of the in-plane electrical resistivity $\rho(T)$ of a MnBi$_2$Te$_4$ crystal was measured from room temperature down to $T = 4.5$ K. It shows an approximately linear decrease of the resistivity upon lowering the temperature, which is characteristic for metals (Figure 4). Below 50 K, $\rho(T)$ deviates from linearity and exhibits a distinct peak-like feature centered at 24.6 K. The local increase followed by a sharp decrease below the peak indicate an enhanced electron scattering in the vicinity of a magnetic phase transition and the freezing-out of scattering due to an onset of long-range magnetic order, respectively.[51] This is consistent with our present magnetic susceptibility and specific heat data. In the antiferromagnetically ordered state the resistivity roughly follows a power law $\rho(T) = \rho(0) + aT^b$ with an exponent $b = 2.2$ and an offset $\rho(0) = 1.1$ m$\Omega$ cm, resulting in a residual-resistivity ratio (RRR) of $\rho(300) / \rho(0) \approx 1.35$. We note that the RRR determined across the 3D magnetic phase transition is of limited applicability because of the associated change in the electronic structure and the freezing-out of scattering. However, despite the tendency of these effects to cause an enhanced RRR, the obtained value is still rather low compared to typical metals, indicating a high impurity scattering rate. The room temperature electrical conductivity of $\sigma = 670$ S cm$^{-1}$ is much higher than the previously reported $\sigma = 39$ S cm$^{-1}$,[25] bringing it much closer to known values of structurally related compounds such as GeBi$_2$Te$_4$ (500 S cm$^{-1}$), SnBi$_2$Te$_4$ (345 S cm$^{-1}$) and PbBi$_2$Te$_4$ (1670 S cm$^{-1}$) at room temperature.[38]

**6. Thermoelectric Characterization.** Earlier reported thermoelectric properties of MnBi$_2$Te$_4$ powders measured under a protective helium atmosphere were not reversible and showed quick deterioration after heating above ca. 277 °C.[25] Moreover, an optical band gap of 0.4 eV was estimated from the diffuse reflectance of a sample that was assumed to be *p*-conducting, judging from a positive Seebeck coefficient.

We used a homogeneous powder sample Mn$_{0.85}$Bi$_{2.1}$Te$_4$ (Figure S14, route #1) for thermoelectric characterization (Figure 5). The Seebeck coefficient as well as electrical and thermal conductivities show a good cyclability, which suggests the absence of a pronounced phase transition or decomposition in the compacted ingot. This is supported by a PXRD pattern obtained from the sample after the measurement of its transport properties. The sample shows only slight decomposition into Bi$_2$Te$_3$ (about 13 wt. %) and MnTe$_2$ (below 1 wt. %, i. e. almost below the detection limit) (Figure S15).

The negative value of the Seebeck coefficient shows electron-dominated charge transport. This is in contrast to structurally related compounds, like GeBi$_2$Te$_4$ and SnBi$_2$Te$_4$, for which the Seebeck coefficients of 102 μV K$^{-1}$ and 64 μV K$^{-1}$, respectively, at room temperature have been reported.[38] On the other hand, it is in line with −59 μV K$^{-1}$ at room temperature for PbBi$_2$Te$_4$.[38] The temperature dependence of the absolute value of the Seebeck coefficient for all these compounds is negative, which is in good accordance with our experiments. The electrical conductivity σ of our polycrystalline samples is about 475 S cm$^{-1}$ at room temperature, in good agreement with the data on a single crystal in the previous section. The negative temperature dependence of σ, i. e. the metallic character, makes the compound stand out when compared to GeBi$_2$Te$_4$ and SnBi$_2$Te$_4$, which both show increasing electrical conductivity in this temperature range. PbBi$_2$Te$_4$, however, exhibits the same metallic characteristics as MnBi$_2$Te$_4$. The thermal conductivity is nearly temperature-independent with only a slight negative trend with increasing temperature. Combined, the enhanced Seebeck coefficient and thermal conductivity overcompensate the lower electrical conductivity at higher temperature, leading to a maximum ZT value of 0.17 at 400 °C.

## CONCLUSION

We have developed the first crystal-growth technique for MnBi$_2$Te$_4$, that grants access to multiple studies on this first antiferromagnetic topological insulator. MnBi$_2$Te$_4$ is an attractive testbed for the realization of various quantum effects, originating from a crossover between non-trivial topology of electronic structure and long-range magnetic ordering.

We have unraveled and solved the synthetic issues that had been limiting the experimental exploration of MnBi$_2$Te$_4$. This compound is stable below its melting point of ca. 600 °C in a narrow temperature window well above room temperature. Since the decomposition of synthesized single crystals and compact polycrystalline MnBi2Te4 is kinetically hindered, samples can be quenched to room temperature. Long-term heating outside the stability range leads to decomposition of powder samples, the decomposition rate likely depending on the grain size. Our crystal structure elucidation using single-crystal X-ray diffraction and electron microscopy unambiguously reveals the ubiquity of antisite defects in both cation positions, which could not be resolved from earlier powder diffraction data.

Magnetization, specific heat, transport, spectroscopic and thermoelectric measurements identify both single crystals and polycrystalline MnBi$_2$Te$_4$ samples as n-type metallic conductors that undergo a transition into a 3D antiferromagnetically ordered phase at about 24 K. We show that the synthesis of MnBi$_2$Te$_4$ can be scaled-up and that bulk material exists at around room temperature and below despite its metastable character in this temperature range. Angle-resolved photoelectron spectroscopy corroborates the presence of a surface state at the $\bar{\Gamma}$ point with a gapped Dirac-cone-like dispersion, which strengthens MnBi$_2$Te$_4$ as a viable candidate for an antiferromagnetic topologically non-trivial material.[32]

Forthcoming endeavors aim at doping of MnBi$_2$Te$_4$ in order to optimize the position of the Fermi energy within the bulk gap. First-principles calculations foresee that stable compounds isostructural to MnBi$_2$Te$_4$ could exist in the Mn–Bi–Se and Mn–Sb–Te systems,[22] thus, substantiating the feasibility of doping or even expansion of this structure family toward new Mn-containing derivatives. First experimental accounts on MnBi$_2$Se$_4$ thin films produced as a capping layer[52] or as a multilayer MnBi$_2$Se$_4$ / Bi$_2$Se$_3$ heterostructure[48] have appeared recently. The layered MnBi$_2$Te$_4$ material is thus envisioned as a progenitor of the new family of functional materials available both as bulk and thin films.

## EXPERIMENTAL SECTION

**Synthesis and Crystal Growth.** Bi$_2$Te$_3$ and α-MnTe were synthesized by annealing stoichiometric mixtures of the elements (bismuth, MERCK, treated at 220 °C in H$_2$-flow; tellurium, MERCK, > 99.9%; manganese, SIGMA-ALDRICH, treated with SiO$_2$ at 900 °C under vacuum). The starting materials were pulverized by ball milling ("Pulverisette 23", Fritsch) for 25 min under argon atmosphere in a glovebox. The homogeneous powder was pressed into pellets with a diameter of 6 mm with an applied pressure of 0.7 GPa and sealed in an evacuated silica ampule. The binaries were used as precursors for crystal-growth experiments. For this purpose, the binary reactants were ground either manually or by ball milling under argon atmosphere in a glovebox and sealed in an evacuated silica ampule.

The choice of binary precursors was governed by the following considerations: 1) a better control of the Te stoichiometry with respect to the highly volatile elemental tellurium; 2) the structural similarity between the reactants and MnBi$_2$Te$_4$ may promote the solid-state reaction; 3) the probability of any reduction-oxidation reaction leading to the formation of MnTe$_2$ and Bi-rich bismuth tellurides may decrease, since the oxidation states of the cations are the same in the reactants and in the product.

Since MnTe melts peritectically at 1151 °C,[53] annealing of the elements at 950 °C for 1 day following Ref. [25] also does not ensure complete homogenization. Both approaches – from the elements and from the binaries – basically start a similar mixture of Bi$_2$Te$_3$ and MnTe on the stage of a long-term annealing at above 500 °C.

The crystal size in our experiments was influenced by cooling rates within the Ostwald–Miers region and various annealing times at subsolidus temperatures. For small single crystals (up to 100 μm) a rate of 6 K h$^{-1}$ was used with an annealing temperature of 590 °C. Larger single crystals were grown with smaller cooling rates (1 K h$^{-1}$) and longer annealing times at the same temperature.

**Thermochemical Analysis.** The samples were analyzed by means of differential scanning calorimetry (DSC) using a Setaram Labsys ATD-DSC device with a k-probe thermocouple (Ni–Cr/Ni–Al; $T_{max}$ = 800 °C) and Al$_2$O$_3$ as a reference compound. Heating and cooling rates of 2 K min$^{-1}$ were employed. Weighted amounts of the manually ground samples were sealed in small evacuated silica ampules and exposed to two consecutive heating–cooling cycles up to the chosen temperature.

**X-Ray Diffraction Experiments.** Single-crystal X-ray diffraction data were collected on a four-circle Kappa APEX II CCD diffractometer (Bruker) with a graphite(002)-monochromator and a CCD-detector at $T$ = 296(2) K. Mo-$K_\alpha$ radiation ($\lambda$ = 71.073 pm) was used. A numerical absorption correction based on an optimized crystal description was applied,[54] and the initial structure solution was performed in JANA2006.[55] The structure was refined in SHELXL against $|F_o|^2$.[56,57] Further details on the crystal structure investigations of $Mn_{0.85(3)}Bi_{2.10(3)}Te_4$ and $MnBi_2Te_4$ can be obtained from the Fachinformationszentrum Karlsruhe, 76344 Eggenstein-Leopoldshafen, Germany (fax, (+49)7247-808-666; E-mail, crysdata@fiz-karlsruhe.de), on quoting the depository numbers CSD-1876907 and CSD-1876906.

Powder X-ray diffraction data were measured using an X'Pert Pro diffractometer (PANalytical) with Bragg–Brentano geometry or a Huber G670 diffractometer with integrated imaging plate detector and read-out system. Both machines operate with a curved Ge(111) monochromator and Cu-$K_{\alpha 1}$ radiation ($\lambda$ = 154.06 pm). Variable divergence slits were used on the X'Pert Pro equipment to keep the illuminated sample area constant. Phase fractions in the powder samples have been estimated by Rietveld refinements of the data; the fundamental parameter approach was applied in the TOPAS package.[58] Preferred orientation of the crystallites was described by spherical harmonics functions. The graphics of the structures were made with the Diamond software.[59]

**Transmission Electron Microscopy.** An FEI Helios NanoLab 600i focused ion beam system was used to cut a TEM-transparent lamella from a $MnBi_2Te_4$ single crystal. SAED patterns were acquired on a TEM FEI Tecnai G2 microscope (200 kV accelerating voltage, $LaB_6$ filament). HR-STEM and EELS investigations have been performed using an FEI Titan3 80–300 double aberration corrected microscope with a GATAN GIF Tridiem 865ER spectrometer, operated at 80 kV accelerating voltage. The convergence semi-angle of the electron beam was set to 21.5 mrad. Due to the weak Mn signal and strong contamination during the spectrum image acquisition, post data treatment using a principal component analysis (PCA) was applied.[60] Six main loadings have been used to reconstruct and denoise the spectrum image.

**Energy-Dispersive X-ray Spectroscopy.** Samples were prepared by embedding small crystals of $Bi_2Te_3$, MnTe and the title compound into synthetic resin (versosit), subsequent polishing to ensure a flat surface and thin carbon coating to provide a conducting surface. EDX spectra were acquired with a high-resolution SEM EVOMA 15 (Zeiss) equipped with a Peltier-cooled Si(Li) detector (Oxford Instruments) employing 30 kV acceleration voltage. Element quantification was obtained from least-square fitting of edge models (Mn-$K$, Te-$L$, Bi-$L$ and Bi-$M$) invoking $k$ factor calibration from the stoichiometric samples of similar composition ($Bi_2Te_3$ and MnTe). To assess systematic errors stemming from different edge and reference choices we included both Bi-$L$ and $M$-edge as well as $Bi_2Te_3$ and MnTe references for Te in our quantification statistics.

**Photoemission and X-ray Absorption Experiments.** $MnBi_2Te_4$(001) surfaces were prepared *in situ* in ultrahigh vacuum (UHV) by cleaving single crystals using a metal post glued on top of the sample.

We performed ARPES experiments on $MnBi_2Te_4$(001) at the Microscopic And Electronic STRucture Observatory (MAESTRO) at beamline 7 of the Advanced Light Source (ALS). Measurements were performed in ultra-high vacuum of lower than $1\cdot 10^{-10}$ mbar for samples cooled down to approximately 18 K. The total energy and momentum resolutions were ca. 20 meV and ca. 0.01 Å$^{-1}$.

XPS and XAS measurements were performed at the ASPHERE III endstation at the Variable Polarization XUV Beamline P04 at PETRA III (DESY), providing a high photon flux over a wide range of photon energies ($h\nu$ = 250–3000 eV). The energy resolution of the XPS measurements was approximately 100 meV.

We performed XLD experiments at the BOREAS beamline of the ALBA synchrotron light source using linearly polarized X-rays. The experiments were carried out at temperatures of $T$ = 2 K and $T$ = 40 K in the total electron yield (TEY) mode. Grazing and normal incidence measurement geometries were achieved by rotating the sample relative to the light beam.

**Magnetic Characterization.** Temperature- and field-dependent magnetic measurements were performed on a $MnBi_2Te_4$ melt ingot (ca. 58 mg) using a Superconducting Quantum Interference Device (SQUID) magnetometer (MPMS®3) from Quantum Design, Inc.

Temperature dependent zero field-cooled and field-cooled warming magnetization measurements were performed at a magnetic field of 1 T in the temperature range of 2–320 K and field-dependent magnetization measurements was done at temperatures 2 K and 300 K.

All samples were mounted with the glue *Duosan* on a silica glass holder and the magnetization data were corrected by the magnetic moment of the glue. The Néel temperature was determined by $\partial(\chi\cdot T)/\partial T$ according to *Fisher*.[61] The critical field of the spin-flop-transition was determined by the derivative of the magnetization with respect to the magnetic field.

**Specific Heat.** Specific heat measurements were performed on several single crystals (ca. 1.6 mg in total) with a preferred orientation (plane (001) perpendicular to $H$) between 1.8 K and 50 K using a heat-pulse relaxation method in a Physical Properties Measurement System (PPMS) from Quantum Design. The heat capacity of the sample holder (addenda) was determined prior to the measurements in order to separate the heat capacity contribution of the sample from the total heat capacity. The transition temperature $T_N$ was determined via an entropy-conserving construction.

**Resistivity Measurements.** The in-plane electrical resistivity of a $MnBi_2Te_4$ single crystal was measured in a home-built setup utilizing a standard four-point geometry in a liquid helium environment. Two cooling-heating cycles from room temperature to 4.5 K yielded reproducible resistivity data with no hysteresis, proving good chemical and mechanical stability of the sample and the contacts applied to its surface. The linear response was confirmed by varying the excitation current in the second cycle, resulting in an identical $\rho(T)$ curve. The electrical current density was assumed homogeneous across the single crystal's cross section. However, due to the layered structure, the current density might be increased close to the sample surface where the electrical field gradient is picked up, resulting in an overestimation of the resistivity. The absolute values of the electrical conductivity given in the results section should therefore be regarded as a lower limit only.

**Thermoelectric Measurements.** Thermal diffusivity measurements were performed under He atmosphere with a Linseis LFA1000 apparatus equipped with an InSb detector. Simultaneous heat loss and finite pulse corrections were corrected using *Dusza*'s model.[62] The values were averaged over five measurement points at each temperature. The samples, polished to a uniform thickness of 1.88 mm, were placed on an aperture of 3 mm diameter. For the κ calculation, diffusivity values were multiplied with the Dulong–Petit heat capacity ($C_p$ = 0.1739 J g$^{-1}$K$^{-1}$) and the density as derived by the weight and the volume determined by *Archimedes*' principle ($\rho$ = 7.409 g cm$^{-3}$). For isostructural materials in the Ge–Sb–Te system, the Dulong–Petit approximation yielded $C_p$ values close to the measured ones.[63] The Seebeck coefficient $S$ and the electrical conductivity σ were measured simultaneously under He atmosphere with a Linseis LSR-3 instrument with NiCr/Ni and Ni contacts and a continuous reverse of the polarity of the thermocouples (bipolar setup). Specimens were cut from the ingots along different directions with dimensions of 7 mm × 2 mm × 2 mm. The errors of S and σ are smaller than 10%; for κ, they are ca. 5%. As a result, the *ZT* values given may exhibit an absolute uncertainty of up to 20%.

## ASSOCIATED CONTENT

**Supporting Information**. Details of powder synthesis, studies of thermal stability and crystal-structure elucidation are available free of charge via the Internet at http://pubs.acs.org.

## AUTHOR INFORMATION

### Corresponding Author

* Email: anna.isaeva@tu-dresden.de

### Author Contributions

The manuscript was written through contributions of all authors. All authors have given approval to the final version of the manuscript.

### Funding Sources

This work was supported by the German Research Foundation (DFG) in the framework of the Special Priority Program (SPP 1666, IS 250/2-1) "Topological Insulators" and project OE 530/5-1, by the ERA-Chemistry Program (RU 766/15-1), and by CRC "Correlated Magnetism: From Frustration to Topology" (SFB 1143), and by CRC "Tocotronics" (SFB 1170).

### Notes

The authors declare no competing financial interest.

## ACKNOWLEDGMENT

We gratefully acknowledge experimental support by the beamline staff at the beamlines BOREAS (Alba), MAESTRO (ALS) and P04 (PETRA III). We thank Mrs. G. Kreutzer (Leibniz-Institute for Solid State and Materials Research, Dresden) for the EDX measurements.

## ABBREVIATIONS

TI topological insulator; DSC differential scanning calorimetry; PXRD powder X-ray diffraction; SCXRD single-crystal X-ray diffraction; SEM scanning electron microscopy; TEM transmission electron microscopy; ED electron diffraction; HRTEM high-resolution transmission electron microscopy.

**Figures**

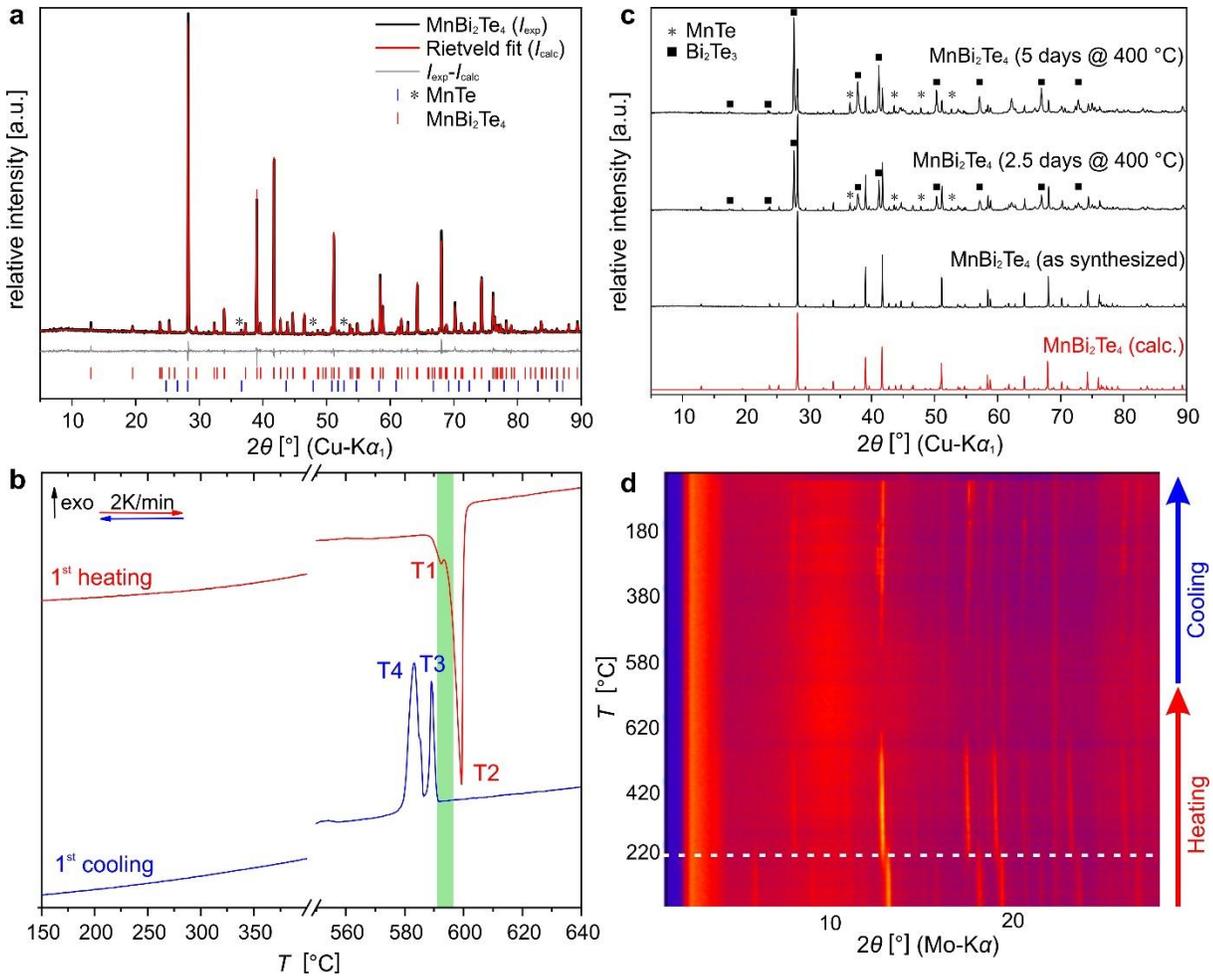

**Fig. 1. Powder synthesis and phase stability of $MnBi_2Te_4$. a)** Refined (red) and experimental (black) PXRD patterns of $MnBi_2Te_4$, sample #2 prepared from a mixture of binaries by a solid-state reaction at 565 °C. The strongest reflections of MnTe admixture are marked by asterisks. The MnTe fraction estimated by a Rietveld refinement is 2 wt. % ($R_{wp}$ = 0.08, $R_{exp}$ = 0.08). **b)** Heating and cooling runs of the DSC experiment on sample #2. The determined Ostwald–Miers region between the peaks onsets is marked in green. **c)** PXRD patterns of the sample #2 as-prepared and subjected to annealing at 400 °C for 2.5 and 5 days, respectively. All annealed samples were water-quenched. **d)** Temperature-programmed PXRD pattern (Mo-$K_\alpha$ radiation) of a fine powder of $Mn_{0.85}Bi_{2.1}Te_4$ heated from room temperature up to 620 °C and cooled back to room temperature at 2 K min$^{-1}$. Decomposition of the ternary phase starting at ca. 220 °C is marked by a white dashed line.

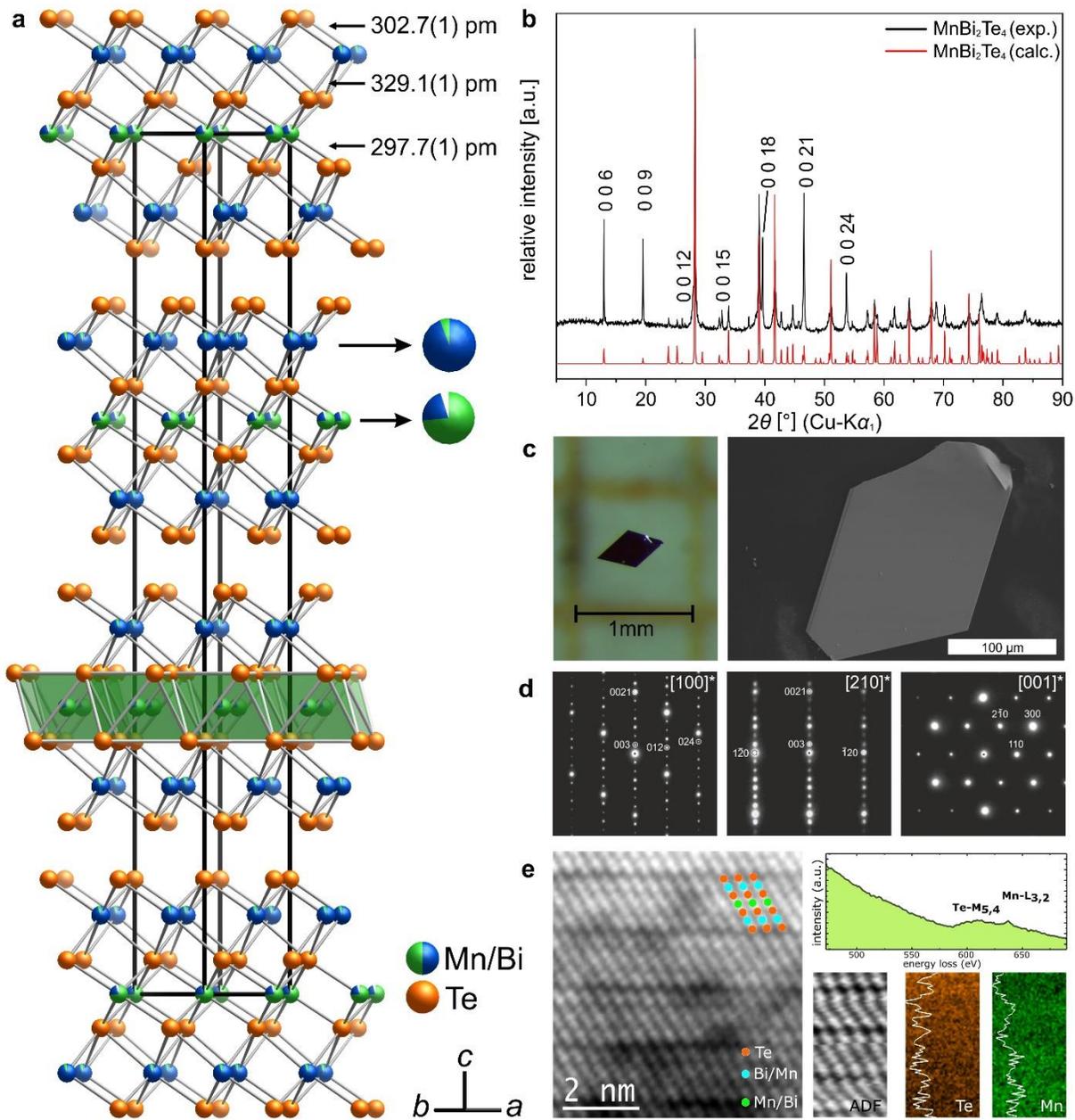

**Fig. 2. Crystal structure and cation intermixing. a)** Crystal structure of the $Mn_{0.85(3)}Bi_{2.10(3)}Te_4$ model with mixed site occupancies and $U_{iso}$ with 99% probability. Selected Mn–Te and Bi–Te interatomic distances are listed. **b)** PXRD of ground crystals prepared via slow cooling (6 K h$^{-1}$) and long-term annealing below the established Ostwald–Miers region. Note strong preferred orientation along [001] in comparison with Figure 1a. **c)** As-grown $MnBi_2Te_4$ crystals used for measurements of physical properties (left) and SCXRD (right). **d)** Main-zone SAED patterns of a $MnBi_2Te_4$ crystallite. Indexing accords with the unit cell found by SCXRD (Table 1, S2). **e)** An HAADF image of a $MnBi_2Te_4$ crystallite. Te (orange), Bi (blue) and Mn (green) columns are color-coded with respect to the ordered crystal structure. The elemental distribution of Mn and Te extracted from the electron energy loss spectrum image using the Te-$M_{5,4}$ and the Mn-$L_{3,2}$ edge of the $MnBi_2Te_4$ crystal. The top panel shows the summed EEL spectra, the bottom panel shows the ADF image as well as the Te-$M$ (orange) and Mn-$K$ (green) distribution. Integrated intensities of the Te-$M$ and the Mn-$L$ edge images are depicted by white lines.

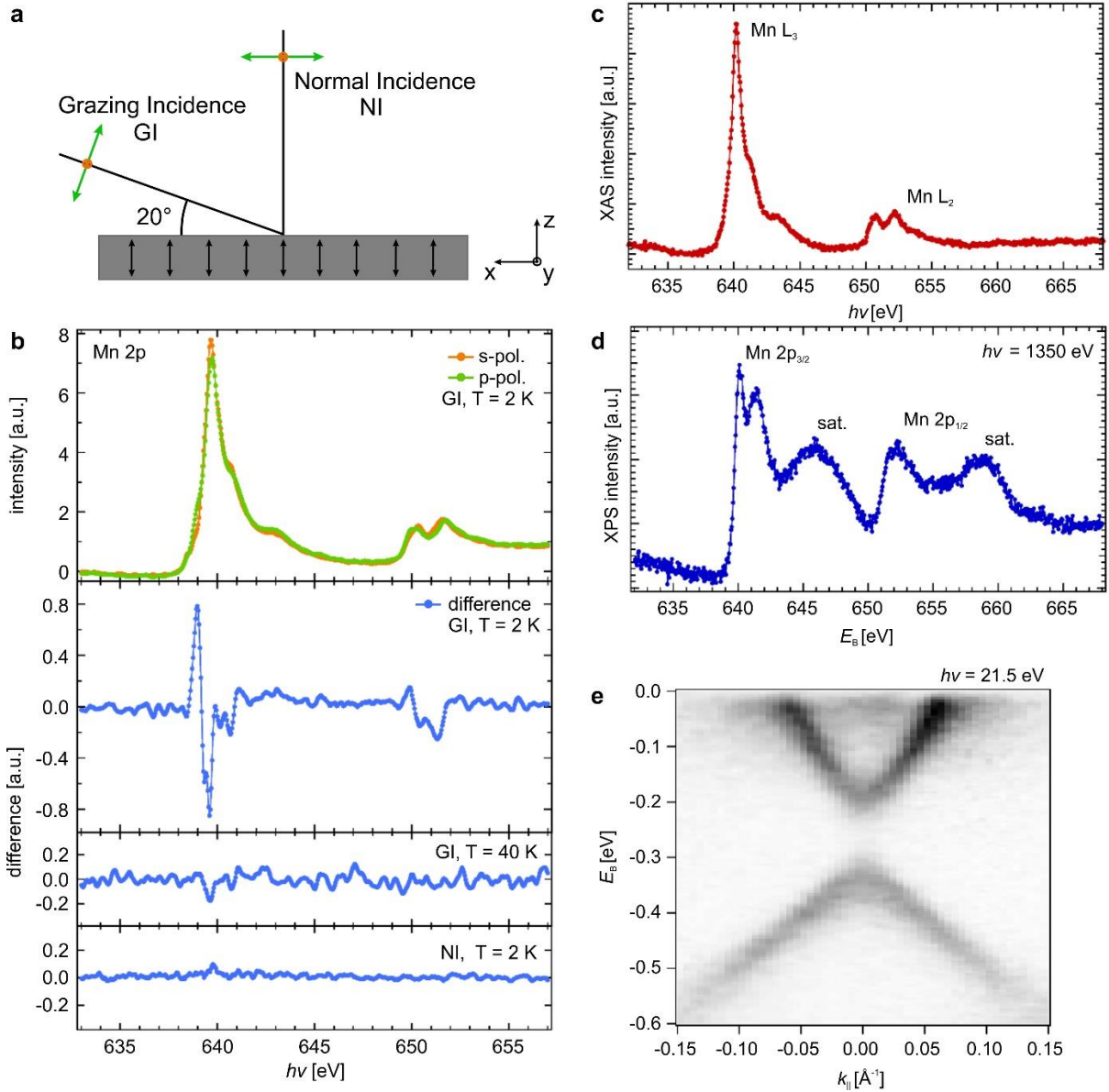

**Fig. 3. XLD and XAS measurements for a MnBi$_2$Te$_4$(001) single crystal. a)** Experimental geometry of XMLD measurements performed in grazing and normal light incidence (GI and NI). **b)** XAS data sets of the Mn $L_{2,3}$ absorption edge obtained using s- and p-polarized light. XMLD signal, i. e. the difference between the two above data sets obtained in GI and NI. **c)** Mn $L_{2,3}$ absorption edge (XAS) for MnBi$_2$Te$_4$ measured in total electron yield (TEY) mode. **d)** XPS Mn 2p spectrum for MnBi$_2$Te$_4$ obtained at $h\nu$ = 1350 eV. **e)** Angle-resolved photoemission (ARPES) data near the Fermi level for MnBi$_2$Te$_4$(001) obtained with p-polarized light at $h\nu$ = 21.5 eV ($T$ = 18 K).

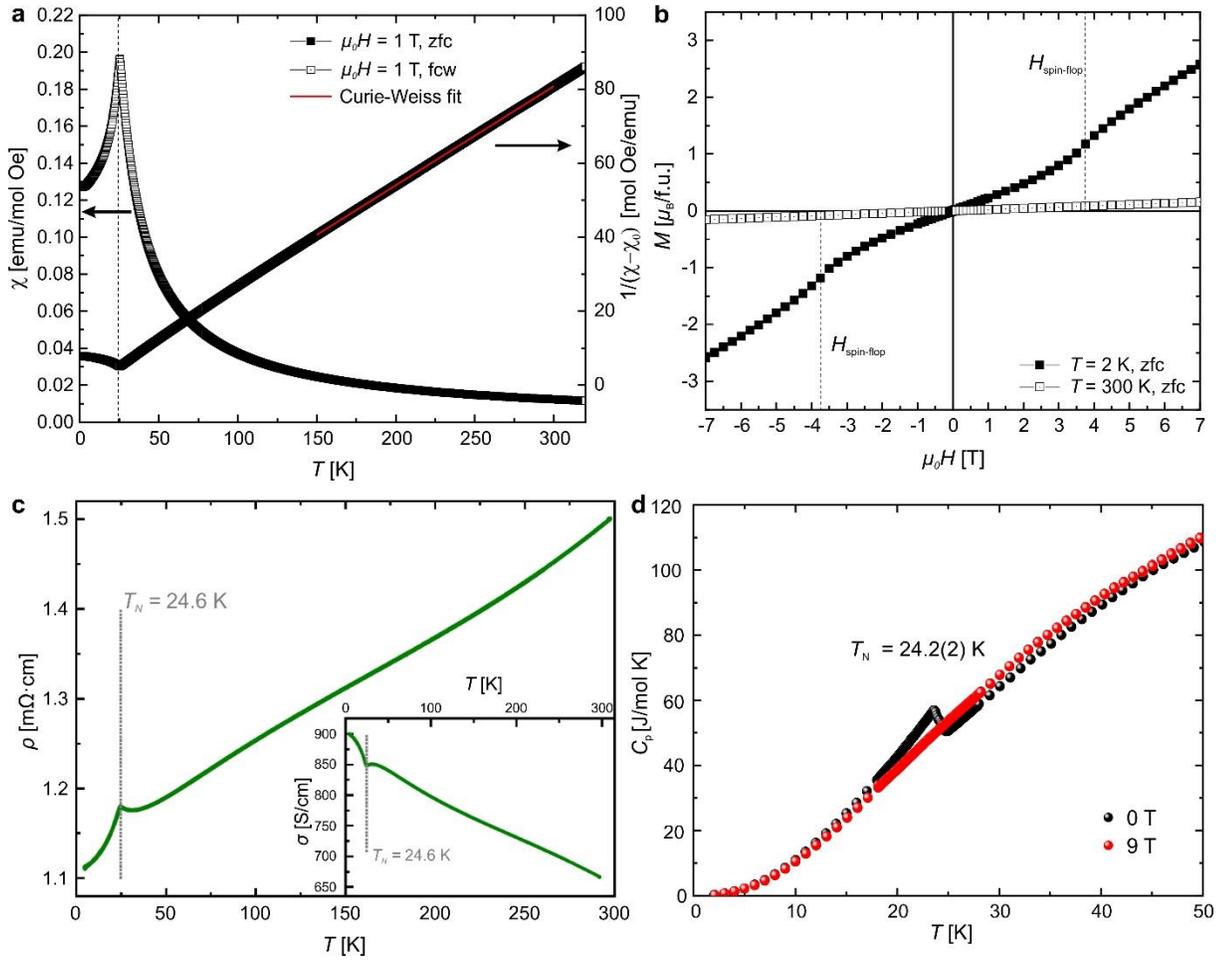

**Fig. 4. Magnetic and transport characterization of polycrystalline (a,b) and single-crystal (c,d) specimens of MnBi$_2$Te$_4$. a)** Temperature-dependent magnetic susceptibility measurements of a melt ingot and its inverse in an external magnetic field of 1 T show an antiferromagnetic transition at $T_N$ = 24.6(5) K. **b)** Field-dependent magnetic moment measurements of a melt ingot. **c)** In-plane electrical resistivity of a single crystal; the inset shows the respective electrical conductivity. **d)** Specific heat measurements of several single crystals.

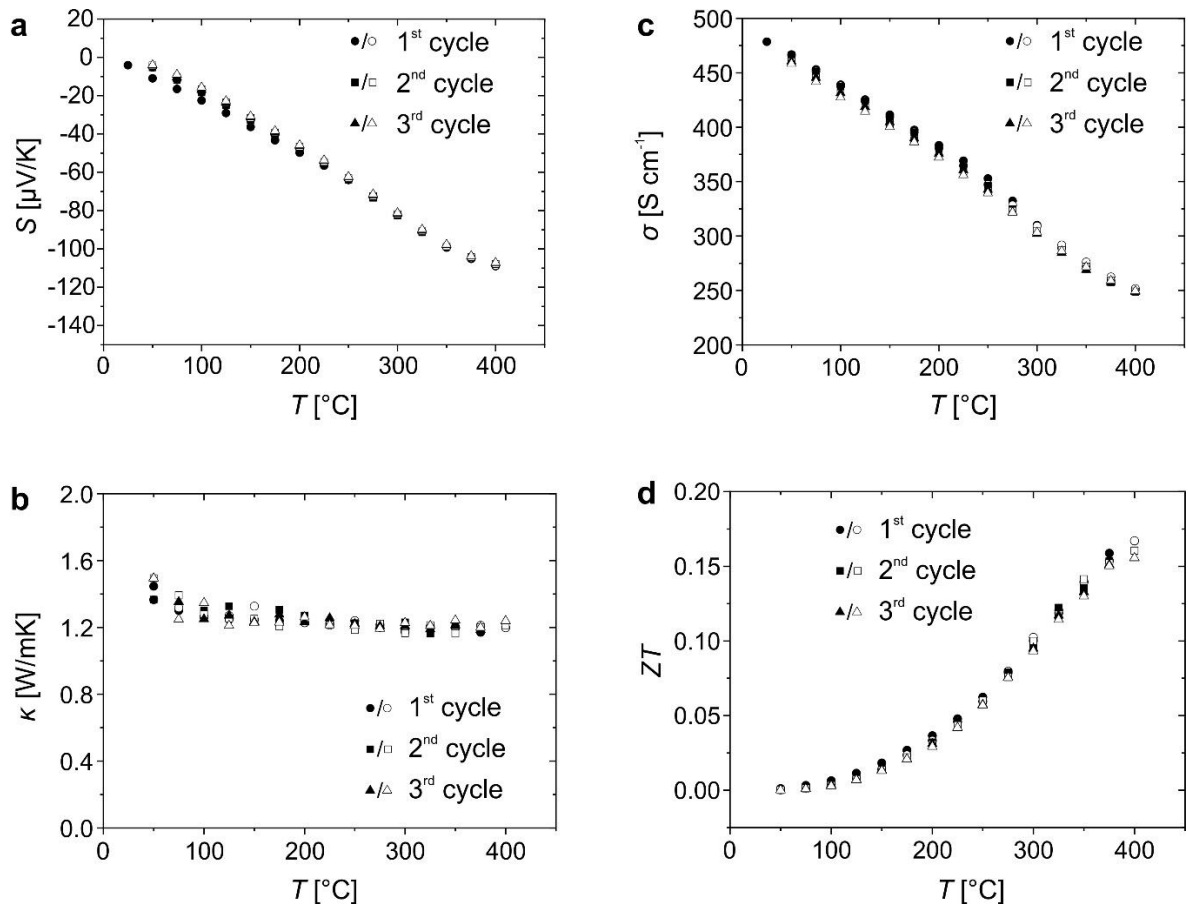

**Fig. 5. Thermoelectric characterization of Mn$_{0.85}$Bi$_{2.1}$Te$_4$ ingot measured in three consecutive cycles. a)** Seebeck coefficient S. **b)** thermal conductivity κ. **c)** electrical conductivity σ. **d)** thermoelectric figure of merit *ZT* as a function of temperature. Filled symbols: heating, open symbols: cooling.

# Tables

**Table 1.** Crystallographic data for $Mn_{0.85(3)}Bi_{2.10(3)}Te_4$, refined with Bi antisite defects and Mn vacancies, from an SCXRD experiment.

| | |
|---|---|
| Crystal system, space group | trigonal, $R\bar{3}m$ (no. 166) |
| Formula units | $Z = 3$ |
| Lattice parameters | $a = 433.14(2)$ pm |
| | $c = 4093.2(2)$ pm |
| | $V = 665.1(1) \times 10^6$ pm³ |
| | $\rho_{calc.} = 7.37$ g cm⁻³ |
| Temperature | 250(2) K |
| Range for data collection; index ranges | $2.98° \leq 2\theta \leq 93.81°$ ($\lambda = 71.073$ pm); $-8 \leq h \leq 8$, $-8 \leq k \leq 8$, $-84 \leq l \leq 83$; |
| Collected reflections | 10795 measured, 852 unique |
| R indices of merging | $R_{int} = 0.079$, $R_\sigma = 0.044$ |
| Structure refinement | Full-matrix least-squares based on $F^2$, anisotropic displacement parameters; |
| Data/restraints/parameters | 852/2/16 |
| Final R indices and goodness-of-fit on $F^2$ | $R1[602\ F_o > 4\sigma(F_o)] = 0.022$ |
| | $wR2$ (all $F_o^2$) = 0.029 |
| | $GooF = 1.039$ |
| Min./max. residual electron density | $-3.62/3.64$ e × 10⁻⁶ pm⁻³ |

**Table 2.** Crystallographic data for $Mn_{0.85(3)}Bi_{2.10(3)}Te_4$, refined with Bi antisite defects and Mn vacancies, from an SCXRD experiment.

| atom | | x | y | z | $U_{eq}$/pm² | SOF |
|---|---|---|---|---|---|---|
| Bi1 | 6c | 1/3 | 2/3 | 0.09173(2) | 182(1) | 0.943(1) |
| Bi2 | 3a | 0 | 0 | 0 | 172(1) | 0.215(1) |
| Mn1 | 6c | 1/3 | 2/3 | 0.09173(2) | 182(1) | 0.057(1) |
| Mn2 | 3a | 0 | 0 | 0 | 172(1) | 0.736(1) |
| Te1 | 6c | 2/3 | 1/3 | 0.03947(2) | 170(1) | 1 |
| Te2 | 6c | 0 | 0 | 0.13337(2) | 169(1) | 1 |